\documentclass[12pt, preprint]{aastex}

\slugcomment{Submitted to {\it The Astrophysical Journal}}

\shorttitle{Modeling the SED and Variability of 3C~66A in 2003 -- 2004}
\shortauthors{Joshi \& B\"ottcher}

\begin{document}

\title{Modeling the Spectral Energy Distribution and Variability of 3C~66A 
during the WEBT campaign of 2003 -- 2004}

\author{M. Joshi\altaffilmark{1} and M. B\"ottcher\altaffilmark{1}}

\altaffiltext{1}{Astrophysical Institute, Department of Physics and Astronomy,
 \\
Ohio University, Athens, OH 45701, USA}

\begin{abstract}
The BL~Lac object 3C~66A was observed in an extensive multiwavelength 
monitoring campaign from July 2003 till April 2004. The spectral energy 
distribution (SED) was measured over the entire electromagnetic spectrum, with
 flux measurements from radio to X-ray frequencies and upper limits in the 
very high energy (VHE) $\gamma$-ray regime. Here, we use a time-dependent 
leptonic jet model to reproduce the SED and optical spectral variability 
observed during our multiwavelength campaign. Our model simulations could 
successfully reproduce the observed SED and optical light curves and predict 
an intrinsic cutoff value for the VHE $\gamma$-ray emission at $\sim$ 4 GeV. 
The effect of the optical depth due to the intergalactic infrared background 
radiation (IIBR) on the peak of the high-energy component of 3C~66A was found 
to be negligible. Also, the presence of a broad line region (BLR) in the case 
of 3C~66A may play an important role in the emission of $\gamma$-ray photons 
when the emission region is very close to the central engine, but further out,
 the production mechanism of hard X-ray and $\gamma$-ray photons becomes 
rapidly dominated by synchrotron self-Compton emission. We further discuss the 
possibility of an observable X-ray spectral variability pattern. The simulated
 results do not predict observable hysteresis patterns in the optical or soft 
X-ray regimes for major flares on multi-day time scales. 
\end{abstract}


\section{Introduction}

Blazars are the most extreme class of Active Galactic Nuclei (AGN) exhibiting 
rapid variability at all wavelengths and a high degree of linear polarization 
in the optical. They have been observed at all wavelengths, from radio through
 VHE $\gamma$-rays and are characterized by non-thermal continuum spectra and 
radio jets with individual components often exhibiting apparent superluminal 
motion. This class of AGNs is comprised of BL Lac objects and flat-spectrum 
radio quasars (FSRQs), which are distinguished primarily on the basis of the 
absence or presence of broad emission lines in their optical spectra.

The broadband spectra of blazars are associated with non-thermal emission and 
exhibit two broad spectral components. The low energy component is due to 
synchrotron emission from non-thermal electrons in a relativistic jet whereas 
the high energy component is attributed either to the Compton upscattering of 
low energy radiation by the synchrotron emitting electrons (for a recent 
review see, e.g., \cite{bo2006}) or the hadronic processes initiated by 
relativistic protons co-accelerated with the electrons \citep[]{mp2001, mp2003}
. Blazars are often known to exhibit variability at all wavelengths, varying 
on time scales from months, to a few days, to even less than an hour in some 
cases. The radio emission of blazars shows variability on a time scale of 
weeks to months whereas the optical emission for some blazars might vary on a 
time scale of around one and a half hours. At X-ray energies, some HBLs exhibit
 characteristic loop features when the photon energy spectral index, $\alpha$,
 is plotted against the X-ray flux. These plots are known as hardness-intensity
 diagrams (HIDs) and the loop structures are called spectral hysteresis. This 
spectral hysteresis can be interpreted as the signature of synchrotron 
radiation, due to the gradual injection and/or acceleration of 
ultrarelativistic electrons in the emitting region and their subsequent 
radiative cooling \citep[]{ki1998, gm1998, ka2000, ku2000, li2000, bc2002}.

3C~66A is classified as a low-frequency peaked (or radio selected) BL~Lac 
object (LBL). The peak of the low-frequency component of LBLs generally lie in
 the IR or optical regime, whereas the high-energy component peak is 
located at several GeV, and the $\gamma$-ray output is typically comparable to
 or slightly higher than the spectral output of the synchrotron component. The
 redshift of 3C~66A has a relatively uncertain determination of $\rm z = 0.444$
 \citep{br2005}. It has exhibited rapid microvariability at optical and near 
infrared in the past and has been suggested as a promising candidate for 
detection by the new generation of atmospheric \v{C}erenkov telescope 
facilities like H.E.S.S., MAGIC, or VERITAS \citep{cg2002}. This object has 
been studied in radio, IR, optical, X-rays and $\gamma$-rays in the past. Its
low-frequency component is known to peak in the IR - UV regime whereas the 
high-frequency component generally peaks at multi MeV - GeV energies. The 
multiwavelength SED and correlated broadband spectral variability behaviour of
 3C~66A have been very poorly understood. For this reason, \cite{bh2005} 
organized an intensive multiwavelength campaign to observe this object from 
July 2003 through April 2004, with the core campaign period being Sept. - Dec.
 2003. 

As described in \cite{bh2005}, the object exhibited several outbursts in the 
optical. The variation was on the order of $\Delta$m $\sim$ 0.3-0.5 over a 
timescale of several days. The minimum variability timsecale of 2 hr provided 
an estimate for the size of the emitting region to be on the order of 
$10^{15}$~cm. The optical flares suggested the presence of an optical spectral
 hysteresis pattern with the B - R hardness peaking several days before the 
R- and B- band flux peaked. The RXTE PCA data indicated a transition between 
the synchrotron and the high-energy component at photon energies of 
$\gtrsim 10$~keV. The broadband SED of 3C~66A suggested that the synchrotron 
component peaked in the optical. In the VHE $\gamma$-ray regime, STACEE 
provided an upper limit at $E_{\gamma} \gtrsim 150$~GeV whereas an upper 
limit at $E_{\gamma} > 390$~GeV resulted from simultaneous Whipple 
observations. 

In this paper, we use a leptonic jet model to reproduce the broadband SED and 
the observed optical spectral variability patterns of 3C~66A and make 
predictions regarding observable X-ray spectral variability patterns and 
$\gamma$-ray emission. In \S \ref{model}, we describe the time-dependent 
leptonic jet model used to reproduce the observed SED and optical spectral 
variability patterns of 3C~66A. The parameters used to simulate the observed 
results are described in \S \ref{parameter}. The modeling results and VHE 
$\gamma$-ray predictions are discussed in \S \ref{results}. We summarize in 
\S \ref{summary}.  

Throughout this paper, we refer to $\alpha$ as the energy spectral index, 
$\rm F_{\nu}$~[Jy]~$\propto \nu^{-\alpha}$. A cosmology with $\Omega_m = 0.3$, 
$\Omega_{\Lambda} = 0.7$, and $\rm H_0 = 70$~km~s$^{-1}$~Mpc$^{-1}$ is used. 
In this cosmology, and using the redshift of $\rm z = 0.444$, the luminosity 
distance of 3C~66A is $\rm d_L = 2.46$~Gpc.

\section{\label{model}Model Description}

The SEDs and optical variability patterns of 3C~66A were modeled using a 
one-zone homogeneous leptonic jet model. The model assumes injection of a 
population of ultrarelativistic non-thermal electrons and positrons into a 
spherical emitting volume (the ``blob'') of comoving radius $R_{b}$ at a 
time-dependent rate. Since the positrons lose equal amount of energy as the 
electrons via the same radiative loss mechanisms so we do not distinguish 
between them throughout the paper. The injected electron population is 
described by a single power law distribution with a particle spectral index q, 
comoving injection density $Q^{\rm inj}_{e}(\gamma;t)$ ($\rm cm^{-3} s^{-1}$) 
and low- and high-energy cutoffs $\gamma_{1}$ and $\gamma_{2}$, respectively, 
such that $Q^{\rm inj}_{e}(\gamma) = Q^{\rm inj}_{0}(t) \gamma^{-q}$ for 
$\gamma_{1}\leq\gamma\leq\gamma_{2}$, where $Q^{\rm inj}_{0}(t)$ is the 
injection function and is given by,

\begin{eqnarray}
\label{1}
Q^{\rm inj}_{0}(t) = \left\{ \begin{array}{ll}
    {L_{\rm inj}(t) \over \rm V^{\prime}_{b}m_{e}c^{2}} {{2 - q} \over {\gamma^{2 - q}_{2} - \gamma^{2 - q}_{1}}} & \textrm{if $q \neq 2$}\\
\\
    L_{\rm inj}(t) \over \rm V^{\prime}_{b}m_{e}c^{2}ln(\gamma_{2}/ \gamma_{1}) & \textrm{if $q = 2$}
   \end{array} \right.
\end{eqnarray}

where $L_{\rm inj}$ specifies the power of the injected pair population and 
$\rm V^{\prime}_{b}$ is the blob volume in the comoving frame.

The randomly oriented magnetic field B has uniform strength throughout the 
blob and is determined by an equipartition parameter 
$\rm e_{B}$ $\equiv$ $\rm u_{B}$/$\rm u_{e}$ (in the comoving frame), where 
$\rm u_{B}$ is the magnetic field energy density and $\rm u_{e}$ is the 
electron energy density. We keep $\rm e_{B}$ constant so that the magnetic 
field value changes according to the evolving electron energy density value as
 determined by equation \ref{2}. The initial injection of the electron 
population into the blob takes place at a height $\rm z_{0}$ above 
the plane of the central accretion disk. The emitting region travels 
relativistically with a speed 
$\rm v/c = \beta_{\Gamma} = (1-1/\Gamma^2)^{1/2}$ along the jet. 
The jet is directed at an angle $\theta_{\rm obs}$ with respect to the line of
 sight. The Doppler boosting of the emission region with respect to the 
observer's frame is determined by the Doppler factor 
$\delta = [\Gamma(1 - \beta_{\Gamma} \cos\theta_{\rm obs})]^{-1}$, where 
$\Gamma$ is the bulk Lorentz factor.

As the emission region propagates in the jet, the electron population inside 
the blob continuously loses its energy due to synchrotron emission, Compton 
upscattering of synchrotron photons (SSC) and/or Compton upscattering of 
external photons (EC). The seed photons for the EC process include the UV soft
 X-ray emission from the disk entering the jet either directly 
\citep[]{dm1992, ds1993} or after getting reprocessed in the BLR or other 
circumnuclear material \citep[]{sr1994, ds1997}. The time-dependent evolution 
of the electron and photon population inside the emission region is governed, 
respectively, by, 

\begin{equation}
\label{2}
{\partial n_{e} (\gamma, t) \over \partial t} = -{\partial \over \partial \gamma} 
\left[\left({d\gamma \over dt}\right)_{loss} n_{e} (\gamma, t)\right] + Q_{e} (\gamma, t) 
- \frac{n_{e} (\gamma, t)}{t_{e,esc}}
\end{equation}

and

\begin{equation}
\label{3}
{\partial n_{ph} (\epsilon, t) \over \partial t} = \dot n_{ph,em} (\epsilon, t) - 
\dot n_{ph,abs} (\epsilon, t) - \frac{n_{ph} (\epsilon, t)}{t_{ph,esc}}
\end{equation}

Here, $(d\gamma/dt)_{\rm loss}$ is the radiative energy loss rate, due to 
synchrotron, SSC and/or EC emission, for the electrons. $Q_{e} (\gamma, t)$ is
 the sum of external injection and intrinsic $\gamma - \gamma$ pair production
 rate and $t_{\rm e,esc}$ is the electron escape time scale. 
$\dot n_{\rm ph,em} (\epsilon, t)$ and $\dot n_{\rm ph,abs} (\epsilon, t)$ are
 the photon emission and absorption rates corresponding to the electrons' 
radiative losses and, $t_{\rm ph,esc} = (3/4)R_{b}/c$ is the photon escape 
timescale. The time-dependent evolution of the electron and photon population 
inside the blob is followed and radiative energy loss rates as well as photon 
emissivities are calculated using the time-dependent radiation transfer code 
of \cite{bc2002}. 

The model only follows the evolution of the emission region out to sub-pc 
scales and as a result only the early phase of $\gamma$-ray production can 
be simulated. Since the radiative cooling is strongly dominant over adiabatic 
cooling during this phase and the emission region is highly optically thick 
out to GHz radio frequencies, the simulated radio flux is well below the 
actual radio data. We do not simulate the phase of the jet components in which
 they are expected to gradually become transparent to radio frequencies as 
that would require the introduction of several additional, poorly constrained 
parameters.

\section{\label{parameter}Model Parameters}

The model independent parameters that were estimated using the SED and
 optical intraday variability measurements \citep[see][]{bh2005} were used to 
develop an initial set of input parameters:

\begin{eqnarray}
\delta &\approx& 15 \hss \cr
R &\approx& 3.3 \times 10^{15} \; {\rm cm} \hss \cr
B &\approx& 2.9 \, \epsilon_B^{2/7} \; {\rm G} \hss \cr
\gamma_1 &\approx& 3.1 \times 10^3 \hss \cr
\gamma_2 &\approx& 1.5 \times 10^5 \hss \cr
p &\approx& 4 \hss 
\label{parameter_summary}
\end{eqnarray}

Here p is the equilibrium spectral index that determines the optical 
synchrotron spectrum and $p = q + 1$ for strongly cooled electrons. The 
initial set of parameters was modified to reproduce the quiescent as well as 
the flaring state of 3C~66A. Approximately 350 simulations were carried out to
 study the effects of variations of various parameters, such as $\gamma_{1}$, 
$\gamma_{2}$, q, B and $\Gamma$, on the resulting broadband spectra and light 
curves. The set of model parameters that provided a satisfactory fit to the 
quiescent state of 3C~66A involved a value of the Doppler factor, 
$\delta = \Gamma = 24$ and a viewing angle of $\theta_{\rm obs} = 2.4^{\rm o}$.
 These parameters were chosen on the basis of VLBA observations that provided 
the limits on the superluminal motion and indicated bending of the jet towards
 the line of sight thus resulting in a smaller viewing angle and a higher 
Doppler boosting of the emission region as compared to the values inferred from
 the superluminal measurements on larger scales \citep[]{jm2005, bh2005}. The 
fitting of the SED both in the quiescent as well as flaring state of 3C~66A was
 carried out such that the simulated quiescent state does not overpredict the 
X-ray photon flux as X-ray photons are expected to be dominated by the flaring
 episodes. On the other hand, the flaring state was simulated such that the 
resulting time-averaged spectrum passes through the observed time-averaged 
optical as well as X-ray data points. This was achieved by varying individual 
parameters, such as, $\gamma_{1}$, $\gamma_{2}$ and q between the values for 
quiescent and flaring states with time profiles as discussed in the next 
section. A value of $\gamma_{1} = 2.1 \times 10^{3}$, 
$\gamma_{2} = 4.5 \times 10^{4}$ and $q = 2.4$ provided a satisfactory fit to 
the flaring state. Also, during our multiwavelength campaign of 2003 - 2004, 
flux upper limits at multi-GeV - TeV energies could be obtained and as a 
result we could get upper limits on the respective parameters governing the EC
 component. The various model parameters used to simulate the two states of 
3C~66A are listed in Table \ref{model_parameters}.
  
\begin{deluxetable}{cccccccccccc}
\tabletypesize{\scriptsize}
\tablecaption{Model Parameters used to reproduce the quiescent and flaring 
state of 3C~66A as shown in Figures \ref{sed_plot190} and \ref{sed_plot344}, 
respectively. 
Note: $L_{\rm inj}$ is the luminosity with which electron population is 
injected into the blob. $\gamma_{1,2}$ are the low- and high-energy cutoffs of 
electron injection spectrum and q is the particle spectral index. Profile 
stands for the flare profile used to reproduce the optical variability pattern,
 $e_{B}$ is the equipartition parameter and magnetic field B is the 
equipartition value. $\Gamma$ is the bulk Lorentz factor, $R_{b}$ is the 
comoving radius of the blob, $\theta_{\rm obs}$ is the viewing angle and 
$\tau_{\rm T, BLR}$ is the radial Thomson depth of the BLR.}
\tablewidth{0pt}
\tablehead{
\colhead{Fit} & \colhead{$L_{inj}$ [$10^{41}$~ergs/s]} & \colhead{$\gamma_{1}$ [$10^{3}$]} & \colhead{$\gamma_{2}$ [$10^{4}$]} & \colhead{q} & \colhead{Profile} & \colhead{$e_{B}$} & \colhead{B [G]} & \colhead{$\Gamma$} & \colhead{$R_{b}$ [$10^{15}$~cm]} & \colhead{$\theta_{obs}$ [deg]} & \colhead{$\tau_{\rm T, BLR}$}
}
\startdata
1  &    2.7     &  1.8   &      3.0   &     3.1    &     --------   &  1    &   2.4   &  24   &  3.59   &   2.4    &    0\\
2  &    8.0     &  2.1   &      4.5   &     2.4    &     Gaussian   &  1    &   2.8   &  24   &  3.59   &   2.4    &    0\\
3  &	8.0	&  2.1   &      4.5   &     2.4    &     Gaussian   &  1    &   2.8   &  24   &  3.59   &   2.4    &    0.3\\
\enddata
\label{model_parameters}
\end{deluxetable}

\begin{figure}
\plotone{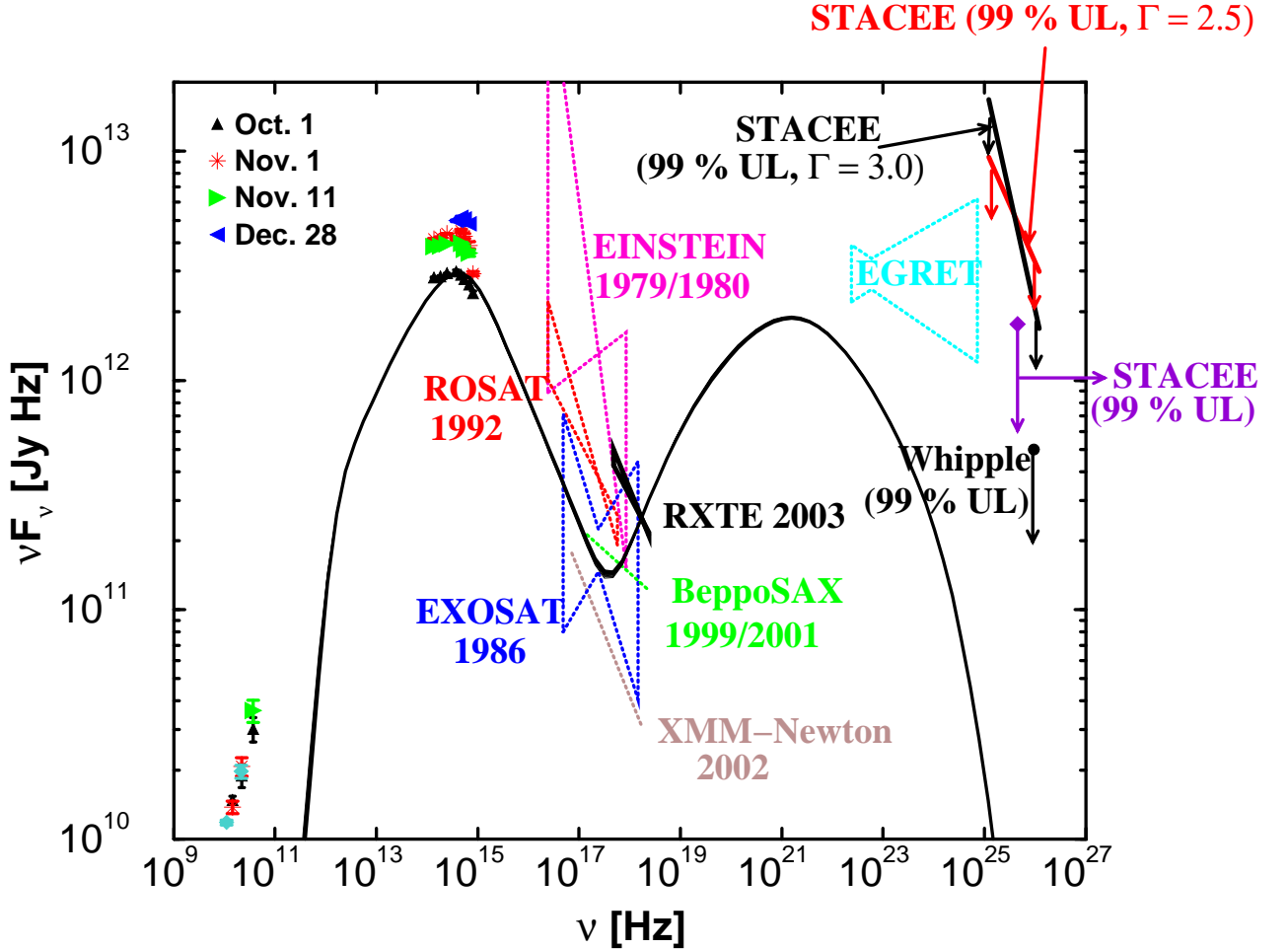}
\caption{Reproduction of the quiescent state of 3C~66A observed around October
 1st 2003. The simulation of this state was carried out using parameters that
do not overpredict the X-ray photon flux. The black colored solid line 
indicates the instantaneous spectrum generated by the simulation after the 
system (blob + injected electron population) attains equilibrium. The 
low-energy component peaks in the optical at $\nu_{\rm syn} \approx 4.8 \times
 10^{14}$~Hz whereas the high-energy SSC component peaks in the MeV regime at 
$\nu_{\rm SSC} \approx 1.6 \times 10^{21}$~Hz. The synchrotron cooling 
timescale in the observer's frame is $\approx$ 1.2 hours, which is on the 
order of observed minimum optical variability timescale of 2 hours. The
 diamond shaped STACEE upper limit is a new addition and is provided by 
\cite{li06}. All data that are indicated by dotted curves are archival data 
and are shown for comparison. The historical average of the 5 EGRET pointings 
is also included to provide a guideline for our simulated VHE emission.}
\label{sed_plot190}
\end{figure}

\begin{figure}
\plotone{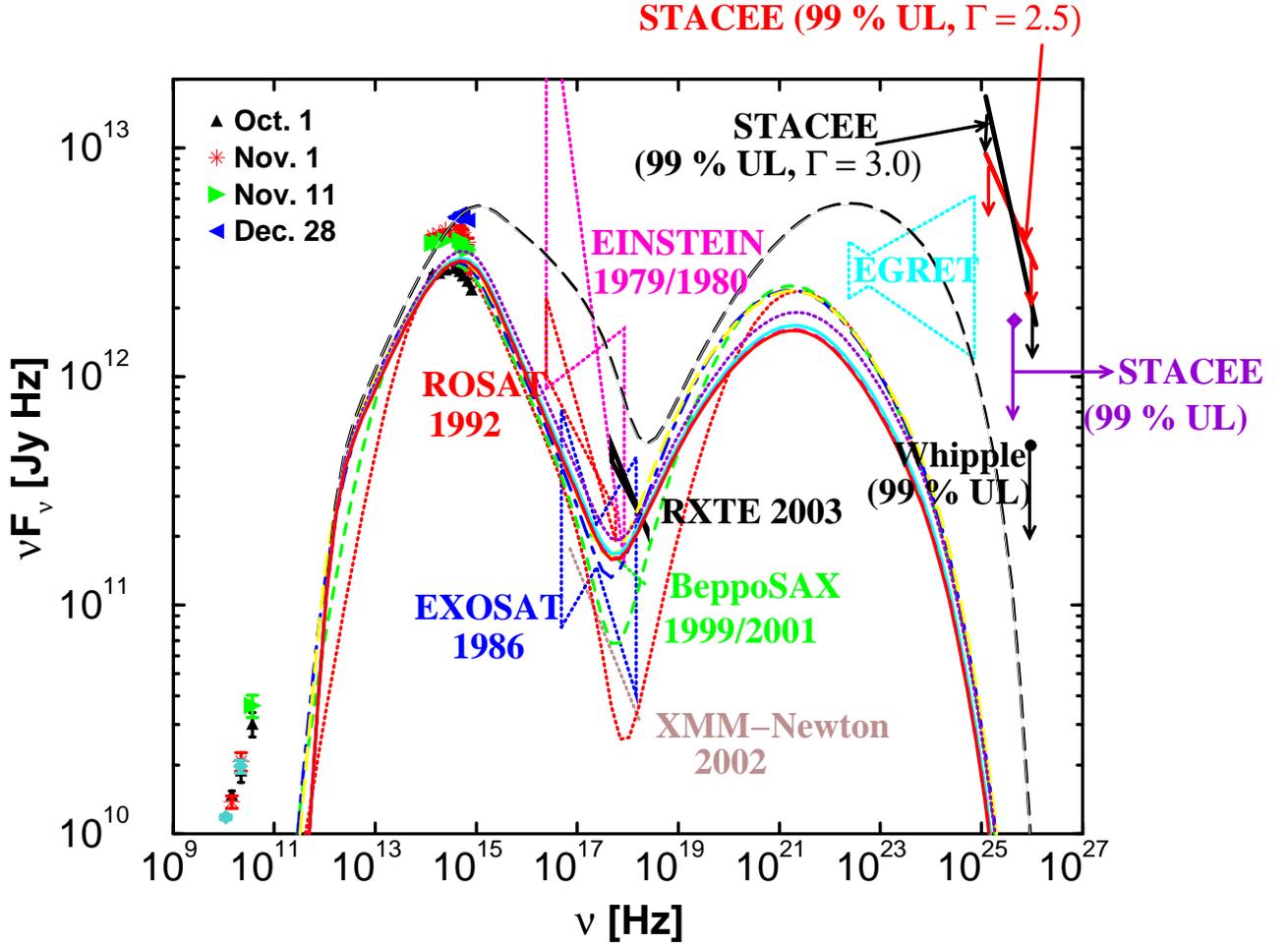}
\caption{Simulation of the flaring state for a generic 10 day flare 
corresponding to the timescale of several major outbursts that were observed 
in the optical regime during our campaign. The various curves show the 
instantaneous spectral energy distribution of 3C~66A at several different 
times in the observer's frame: black (red in the online version) dotted line 
($\sim$ 5th hour), gray (green) dashed line ($\sim$ 8th hour), black (blue) 
dot-dashed line ($\sim$ 14th hour), gray (yellow) long-dashed line 
($\sim$ 20th hour), long-dashed black line ($\sim$ 8th day, highest state 
attained by the system during flaring), gray solid line ($\sim$ 9th day), 
dotted black (violet) line ($\sim$ 16th day), gray (cyan) colored solid line 
($\sim$ 18th day), dashed black (magenta) colored line ($\sim$ 20th day) and 
black (red) solid line ($\sim$ 22nd day, equilibrium state reached by the 
system after the flaring episode is over). The synchrotron component of the 
flaring state peaks at $\nu_{\rm syn} \approx 1.1 \times 10^{15}$~Hz and the 
SSC component peaks at $\nu_{\rm SSC} \approx 2.7 \times 10^{22}$~Hz. The SSC 
component of this state cuts off at $\nu_{\rm SSC, cutoff} 
\approx 2.3 \times 10^{24}$~Hz. The synchrotron cooling timescale in the 
optical regime is $\approx$ 37 minutes for the flaring state.}
\label{sed_plot344}
\end{figure}

Figures \ref{sed_plot190} and \ref{sed_plot344} respectively show the 
reproduction of the SED of 3C~66A, for both the quiescent and flaring state 
observed during the campaign period. The quiescent state is a reproduction of 
the state observed around 1st October 2003 whereas the flaring state is the 
reproduction of a generic 10 day flaring period corresponding to the timescale
 of several of the major outbursts that were observed during the campaign. The
 simulated time-averaged spectrum of 3C~66A in the flaring state is shown in 
Figure \ref{sed_timeav_344}. The simulations, corresponding to fits 1 and 2 of
 Table \ref{model_parameters}, were carried out for a pure SSC emission 
process by artificially setting $L_{D} = 0$, where $L_{D}$ is the bolometric 
disk luminosity. Fit 3 of Table \ref{model_parameters} refers to an EC+SSC 
case with $L_{D} = 1.0 \times 10^{45}~{\rm ergs~s^{-1}}$ and is shown in 
Figure \ref{sed_plot348}. The value of $L_{D}$ was chosen such that it is more
 than the value of the jet luminosity used in the simulations and at the
same time does not produce a blue bump in the simulated SED. In order to 
assess the possible effect of EC emission in 3C66A, an upper limit to the 
optical depth of the BLR was first determined using XSTAR, which returns 
the ionization balance and temperature, opacity, and emitted line 
($H_{\alpha}$, $H_{\beta}$) and continuum fluxes. The BLR was modeled as a 
spherical shell with $r_{\rm BLR, in} = 0.045$~pc and 
$r_{\rm BLR, out} = 0.050$~pc, where $r_{\rm BLR, in}$ and $r_{\rm BLR, out}$ 
stand for the inner and outer radii of the broad line region. A Thomson optical
 depth of 0.3 for the BLR was chosen as a reasonable upper limit such that the
 line emission is weak enough or absent to be consistent with the observed 
featureless continuum.  

\section{\label{results}Results and Discussion}

\begin{figure}
\plotone{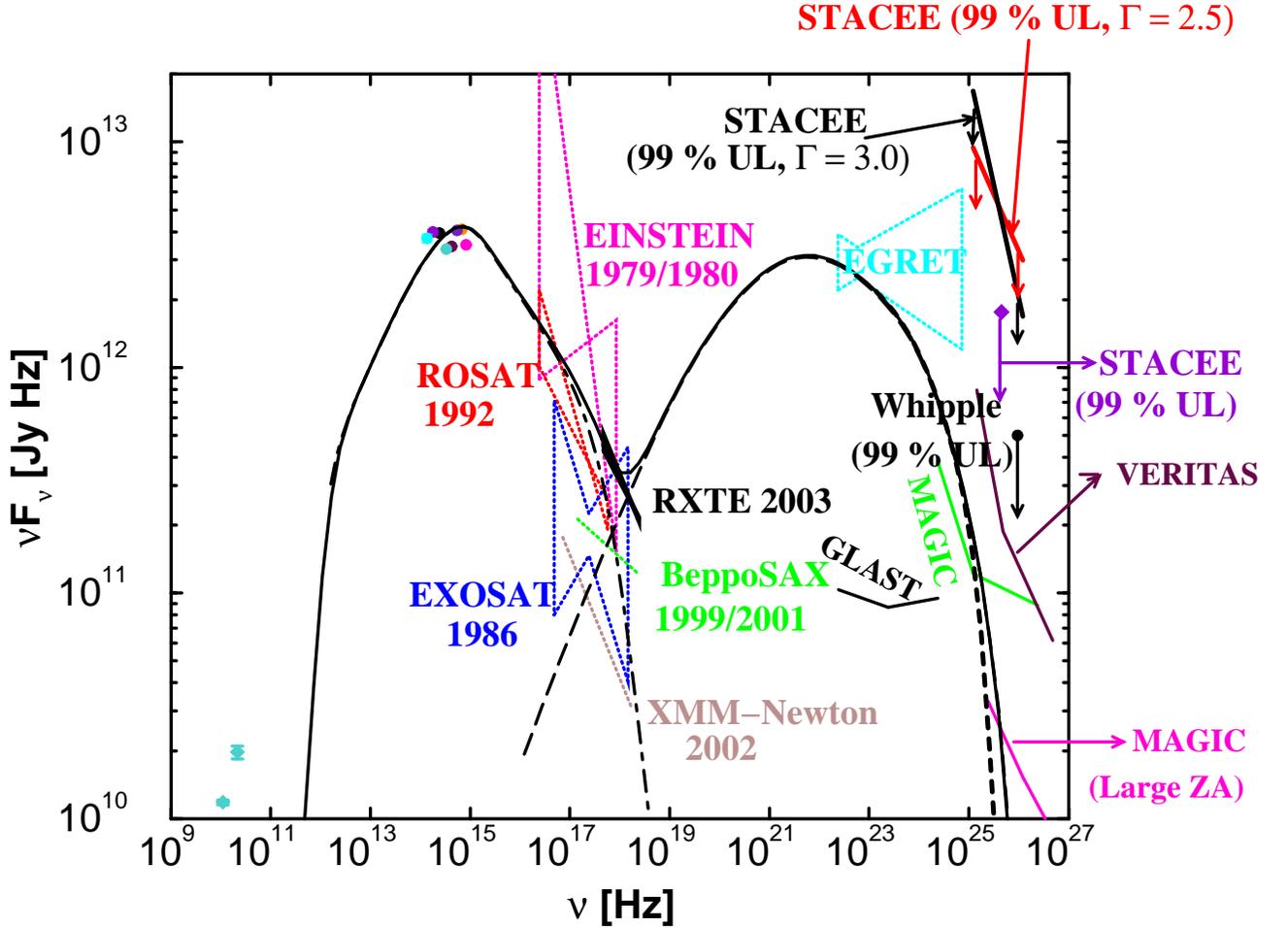}
\caption{Time-averaged spectral energy distribution of 3C~66A for a period of 
23 days around a flare as shown in Figure \ref{sed_plot344}. The filled 
black (colored in the online version) circles are the time-averaged 
optical and IR data points for the entire campaign period and the ``RXTE 2003''
 denotes the time-averaged X-ray data points. The dot-dashed black line is the
 contribution from the synchrotron component only whereas the long-dashed black
 line indicates the contribution of the SSC component only. The time-averaged 
synchrotron component peaks at $\nu_{\rm syn} \approx 7.2 \times 10^{14}$~Hz 
whereas the time-averaged SSC component peaks at 
$\nu_{\rm SSC} \approx 5.3 \times 10^{21}$~Hz. The synchrotron component cuts 
off near 7 keV whereas the SSC component cuts off at $\sim$ 4 GeV. The black 
colored dashed line indicates the attenuation due to the optical depth at VHE 
energies. The $\gamma\gamma$ absorption effect becomes significant at 
$\sim$ 200 GeV. The black (green, maroon and magenta) lines indicate 
the sensitivity limits for an observation time of 50 hours for MAGIC, VERITAS 
and MAGIC (Large Zenith Angle) and for GLAST for an observation time of 1 
month.}
\label{sed_timeav_344}
\end{figure}

As can be seen in Figure \ref{sed_timeav_344}, the time-averaged simulated 
spectrum passes through the time-averaged optical data points whereas the high
 energy end of the synchrotron component passes through the time averaged 
X-ray data indicating the dominance of synchrotron emission in the production 
of such photons in case of flaring. For X-ray photons with energy beyond 
10-12 keV, the data is less reliable due to low count rates and possible source
 confusion with 3C~66B. The spectral upturn at $\geq$ 7 keV occurs due to the 
presence of the SSC component in the simulation. The presence of this component
 cannot be suppressed because in order to suppress it the population of seed 
photons would have to be diluted, which can be done by increasing the size of 
the emission region. But the size of the emission region cannot be increased 
any further due to the strict constraint on the maximum size of the blob that 
comes from the observed minimum variability timescale in the optical region, 
which is 2 hrs. Hence, the emission region size cannot exceed 
$3.6 \times 10^{15}~(D/24)$~cm. Thus, our model suggests that the harder X-ray
 photons come from the SSC and not the synchrotron mechanism with the expected
 spectral hardening taking place at $\sim$ 7 keV. The high energy component, 
due to the SSC emission, for the time-averaged spectrum (see Figure 
\ref{sed_timeav_344}) cuts off at $\sim 1.0 \times 10^{24}$~Hz or 4 GeV. From 
the simulated level of VHE emission we predict that the object is well within 
the observational range of MAGIC, VERITAS and, especially, GLAST (see Figure 
\ref{sed_timeav_344}) whose sensitivity limit is 50 times lower than that of 
EGRET at 100 MeV and even more at higher energies and its two year limit for 
source detection in an all-sky survey is 
$1.6 \times 10^{-9}~{\rm photons~cm^{-2}~s^{-1}}$ (at energies $> 100$~ MeV). 
Thus it will be possible to extract the spectral and variability information
 for this object at such high energies in future observations.

\begin{figure}
\plotone{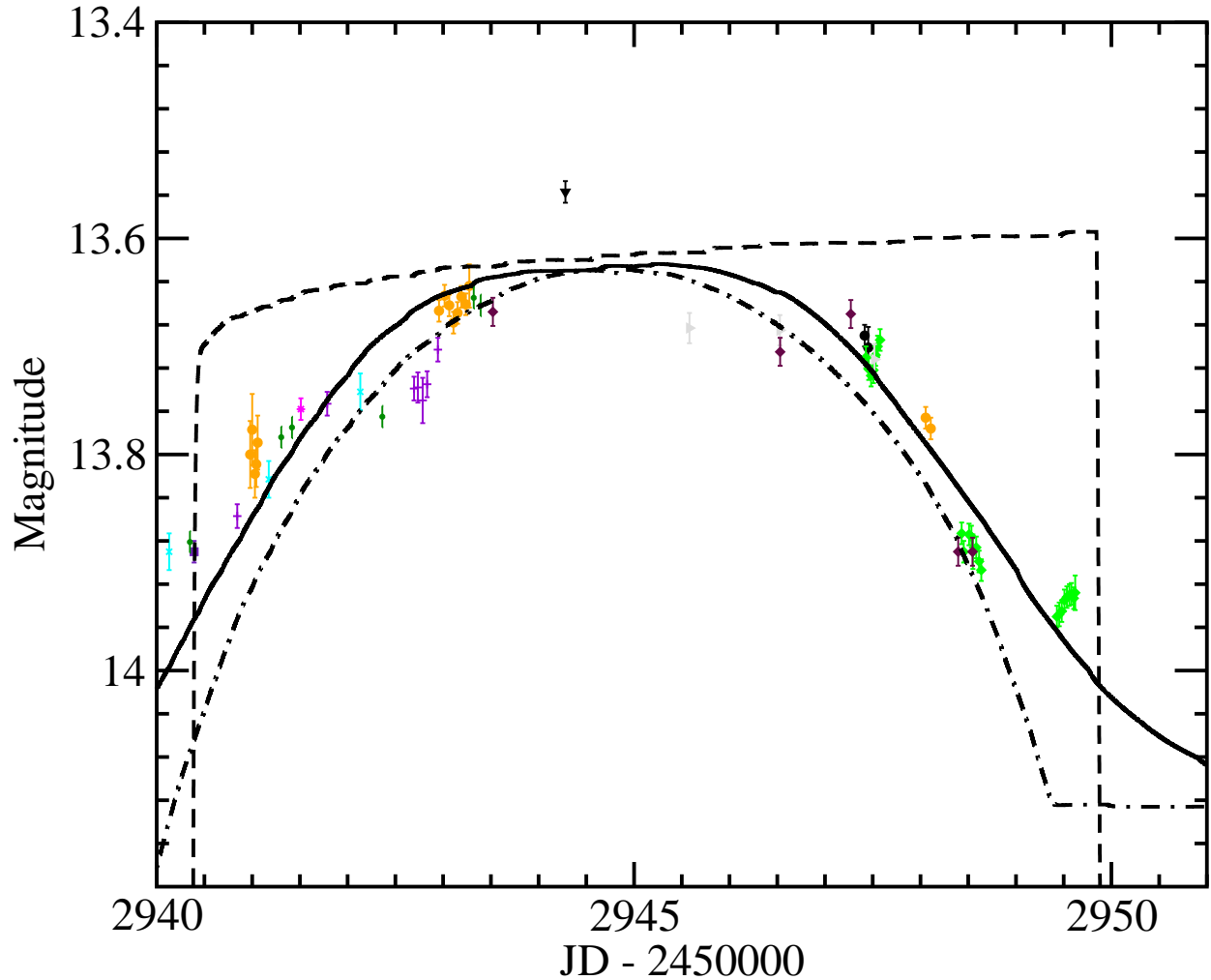}
\caption{The simulated lightcurves for various flaring profiles that have been
 superimposed on the observed R-band lightcurve (see Figure 7 of \cite{bh2005})
 for an outburst on $\sim$ November 1st 2003. The solid black line denotes a 
flaring profile that is Gaussian in time as used for the flare in Figure 
\ref{sed_plot344}, the dash-dotted black line is a trianglular flaring profile
 whereas the dashed black line is a flaring profile that is top-hat in time. As
 can be seen, the Gaussian flaring profile closely matches the width as well 
as the profile of the observed flare.}
\label{flare_profile}
\end{figure}

Flaring above the quiescent state of 3C~66A was reproduced using a 
flaring profile for the electron injection power ($L_{\rm inj}(t)$) 
that was Gaussian in time (see Figure \ref{flare_profile}):

\begin{equation}
\label{5}
L_{\rm inj}(t) = L_{\rm inj}^{\rm qu}(t) + {(L_{\rm inj}^{\rm fl} - L_{\rm inj}^{\rm qu}) 
\over \exp{\left[ (\rm z - r_{c})^2 \over 2 \sigma^{2}\right]}}
\end{equation}

Here, qu and fl stand for the quiescent and flaring state respectively,
 z determines the position of the emission region in the jet at time t, 
$\rm r_{c}$ indicates the position of the center of the simulated 
flare and $\sigma$ stands for the Gaussian width of the flare.

The rest of the parameters such as $\gamma_{1}$ and $\gamma_{2}$ and q were 
also changed accordingly. In order to simulate the observed optical flare, the
 system was first allowed to come to an equilibrium and after the equilibrium 
was set up the flare was introduced with a Gaussian width, $\sigma$ 
corresponding to 14 days in the observer's frame. Although the flare was 
introduced in order to simulate the observed major optical outbursts lasting 
for 10 days, the choice of 14 days for the Gaussian width was made such that 
the width of the simulated flare matches that of the observed flare, 
$\rm r_{c}$ was adjusted such that the centre of the simulated flare aligns 
with that of the observed one and the value of $L_{\rm inj}$ was varied such 
that the peak of the simulated flare matches that of the observed one.

The observed lightcurves did not agree well with a flaring profile that was 
top-hat or triangular in time as can be seen in the figure. The presence of a 
flare that is Gaussian in time might represent an initial injection of 
particles into the emission region at the base of the jet. The particles 
slowly get accelerated as a shock wave ploughs through the region and finally 
dies out in time. Crucial information on the dominant acceleration mechanism 
comes from the change in the shape of the particle injection spectral index 
with time, which might also indicate a possible change in the B-field 
orientation. According to the current understanding of acceleration mechanisms,
 parallel shocks generally produce electron spectra of 
$Q_{e}(\gamma)\propto \gamma^{-q}$ with $2.2\la q\la 2.3$ 
\citep[]{ac2001, ga1999}, whereas oblique shocks produce much softer injection
 spectral indices. On the other hand, 2nd order Fermi acceleration behind the 
shock front might give rise to a harder injection index of the order of 
$q \sim 1$ or beyond \citep{vv2005}. In order to reproduce the flaring state, 
the simulation first starts out in the quiescent state with quiescent state 
parameters and then the value of these parameters is changed to the flaring 
state parameters as the flaring is introduced in the simulation. Since, the 
value of q, in our simulations, changes from 3.1 (quiescent state) to 2.4 
(flaring state) it might indicate a possible change in the orientation of the 
B-field from oblique to parallel during the flaring episode or an interplay 
between the 1st and 2nd order Fermi acceleration thereby making the particle 
spectra harder. The contribution from such acceleration mechanisms and the 
shear acceleration \citep{rd2004} might play an important role in accelerating
 the particles to higher energies. 

\begin{figure}
\plotone{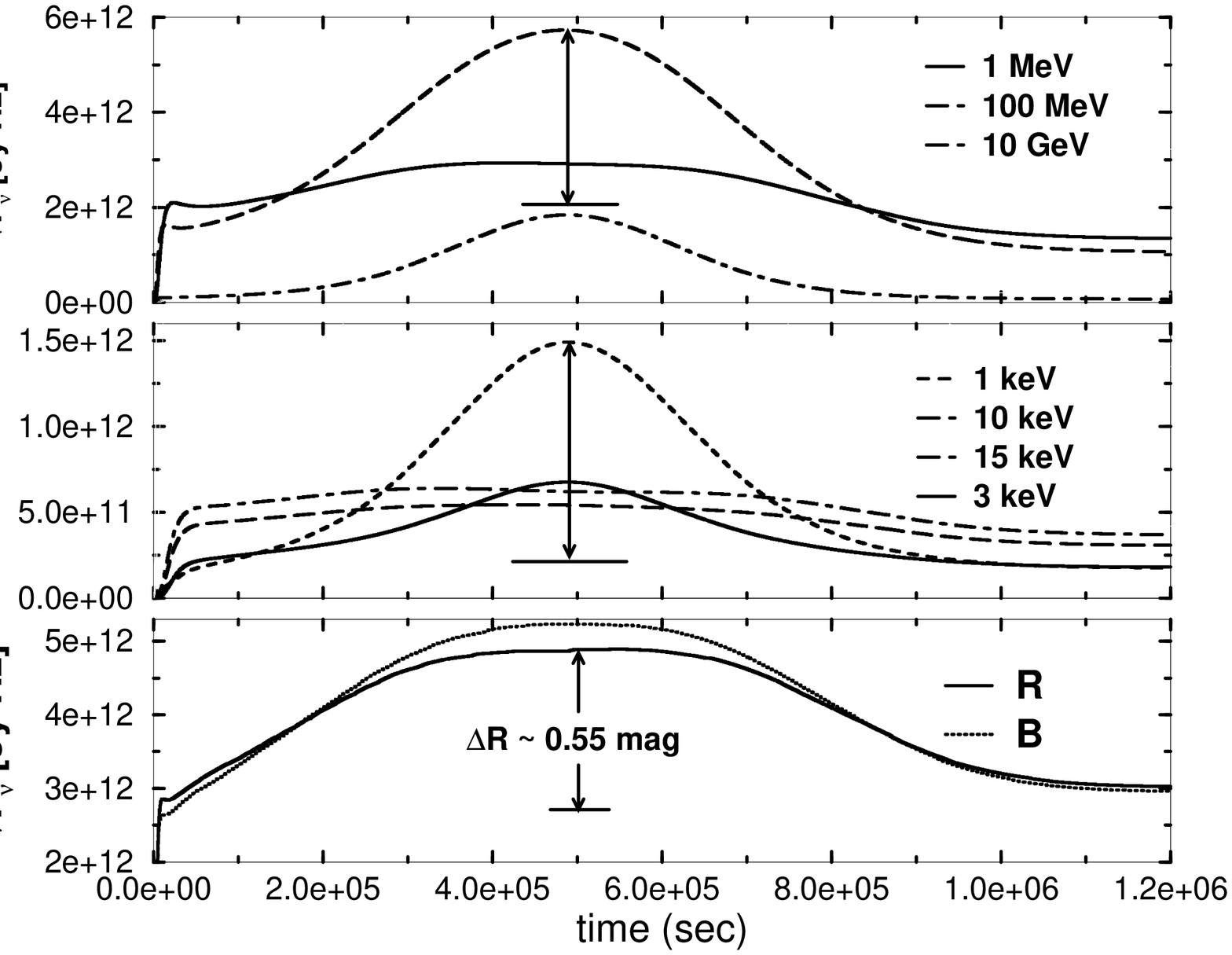}
\caption{Simulated lightcurves for the optical, X-rays and $\gamma$-ray energy
regimes shown in the three panels respectively. The simulated variability in 
the R band is $\approx$ 0.55 mag as indicated by the arrows. The B band, 
denoted by the black dotted line exhibits a higher variability of $\approx$ 
0.7 mag, in the simulation, than that in the R band, which is consistent with 
our observations. The simulated lightcurve at 1 keV is indicated by a black 
dashed curve and exhibits an amplitude variation of 
$\approx 1.4 \times 10^{12}$~Jy Hz. The 3, 10 and 15 keV lightcurves, denoted 
by the black solid line, black long-dashed line and the black dot-dashed 
curve, respectively, on the other hand do not exhibit much variability. In the
VHE regime, the 1 MeV lightcurve is denoted by a black solid line. The 100 MeV
 lightcurve is indicated by a black long-dashed curve and the simulated 
variability amplitude in this energy regime is on the order of $10^{12}$~Jy Hz.
 The black dot-dashed line indicates the lightcurve at 10 GeV.}   
\label{sed_344_lightcurve}
\end{figure}

\begin{figure}  
\plotone{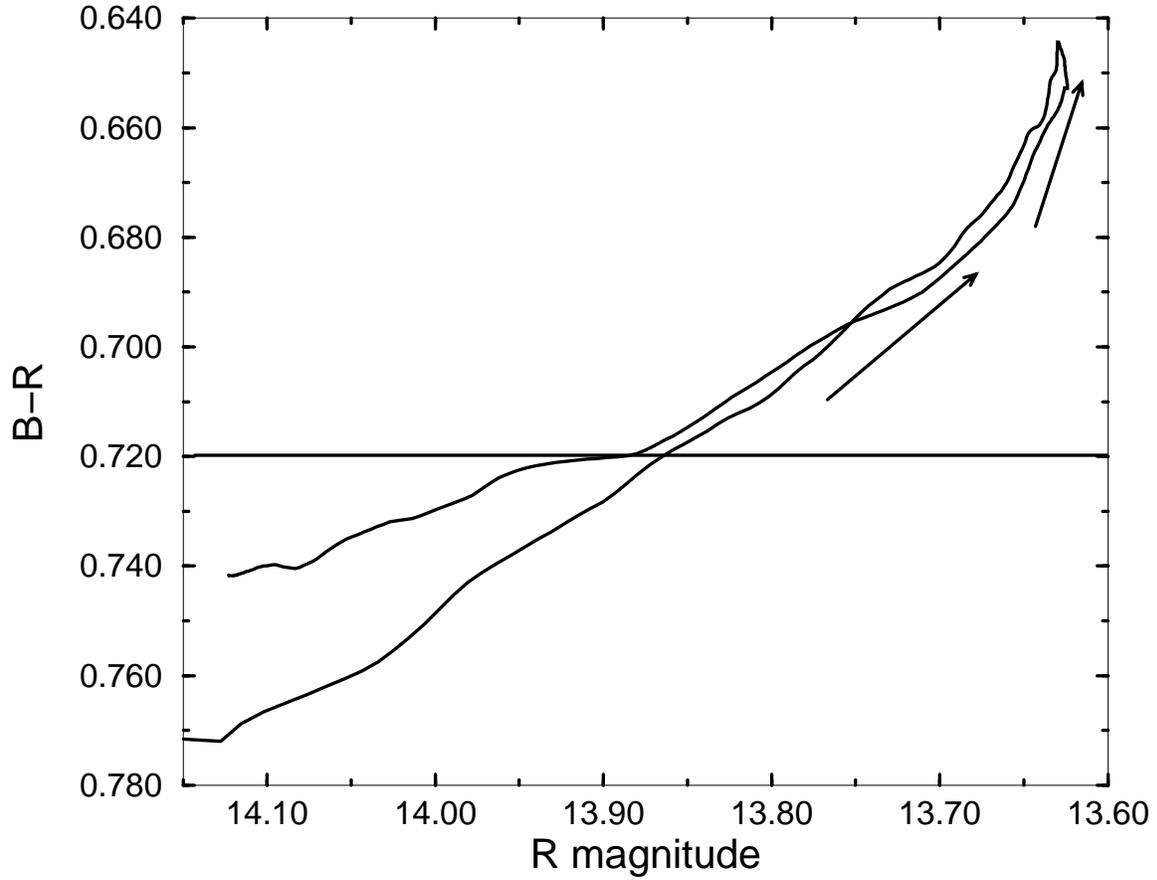}
\caption{The simulated hardness-intensity diagram indicates a positive 
correlation between R- and B-band for an outburst lasting for $\sim$~10 days. 
The object becomes brighter in R and harder in B-R as shown by the arrows. The
spectral upturn takes place at B-R $\approx$ 0.72 mag where the flux in B 
equals that in R (corresponding to $\alpha_{\rm BR} = 0$).}
\label{B-RvsR}
\end{figure}

The simulated optical variability in the R band (0.55 mag) matches the observed
 value (0.3-0.5 mag) for a 10 day period outburst. The predicted 
variability in B is more than that of R by $\sim 0.15$ mag as also observed, 
which indicates that the spectrum is becoming harder (see Figure 
\ref{sed_344_lightcurve}) with the spectral upturn occuring at B-R $\approx$
0.72 mag as shown in Figure \ref{B-RvsR}. Figure \ref{B-RvsR} is a hardness 
intensity graph that shows that the object follows a positive correlation of 
becoming harder in B-R while getting brighter in both the bands during the 
10-day flare simulated in Figure \ref{sed_plot344}. This agrees well with the 
observed optical variability pattern. In this study, we are not 
addressing the variability that was observed on intraday timescales as that 
analysis would open up an even larger parameter space, which cannot be 
reasonably well constrained without any variability information in the X-ray 
regime.    
 
The flare declines faster as compared to the time taken by the flare to rise. 
This might indicate that the particles' synchrotron cooling timescale is less 
than or equal to the light crossing time. 

\begin{equation}
\label{6}
\tau_{\rm cool, sy}^{\rm obs} \approx 2.8 \times 10^3 \, \left( {\delta \over 15} \right)^{-1/2}
\, \left( {B \over 2.9 \, {\rm G}} \right)^{-3/2} \, \nu_{15}^{-1/2} \; {\rm s}
\label{tau_sy}
\end{equation}

We can calculate the observed synchrotron cooling timescale, 
$\tau_{\rm cool, syn}^{\rm obs}$ in the optical regime from equation 
\ref{tau_sy} \citep{bh2005} using $\delta = 24$, $B = 2.4~G$ and 
$\nu_{15} = 0.48$ for the quiescent state and $B = 2.8~G$ and $\nu_{15} = 1.1$
 for the flaring state (see Figures \ref{sed_plot190} and \ref{sed_plot344}), 
where $\nu_{15}$ is the characteristic synchrotron frequency in units of 
$10^{15}$~Hz. This yields a value of 
$\tau_{\rm cool, sy}^{\rm obs} \sim 1.2$~hours for the quiescent state whereas
 for the flaring state it reduces to 37 minutes. The observed minimum 
variability timescale of $\sim$~2 hours might therefore correspond to the 
observed dynamical timescale, where 

\begin{equation}
\label{7}
\tau_{\rm dyn}^{\rm obs} \approx \left( {R_{b} \over c} \right) \, \left( {{1+z} \over D} \right).
\label{tau_dyn}
\end{equation} 

This implies that it takes time to build up the electron population in the 
emission region through flaring but once built up the electrons lose their 
energy efficiently to produce synchrotron photons. This can be used to 
constrain the value of the magnetic field in the jet, which has been allowed 
to evolve in time keeping $e_{B} = 1$ and has an average value of 2.4 Gauss in 
the simulated quiescent state and 2.8 Gauss in the simulated flaring period. 

The crossover of X-ray lightcurves, in our simulations, is a result of the 
dominance of the SSC component in hard X-rays (see Figure 
\ref{sed_344_lightcurve}). The lightcurve of soft X-ray photons of energy 1 
keV exhibits a greater variability of $\sim 1.4 \times 10^{12}$~Jy Hz in its 
flux as compared to their optical counterpart. This is expected because the 
soft X-ray photons, during the flaring episode, are produced from synchrotron 
emission of electrons that are accelerated to very high energies and as a 
result have a very short cooling timescale and thus greater variability. In 
case of hard X-rays no significant variability is predicted. This is because 
such photons are produced from Compton upscattering of synchrotron photons off
 the low-energy electrons and as a result the cooling timescale is much longer
 as compared to the cooling timescale of their soft X-ray and optical 
counterparts. Hence, the variability information gets washed out. The predicted
 X-ray spectral variability pattern of large variability in the low X-ray 
energy band and negligible variability in the high X-ray energy band is 
similar to what has also been observed in BL Lacertae on several occasions 
\citep[see for e.g.,][]{ra2003, ra2002}. 
 
\begin{figure}
\plotone{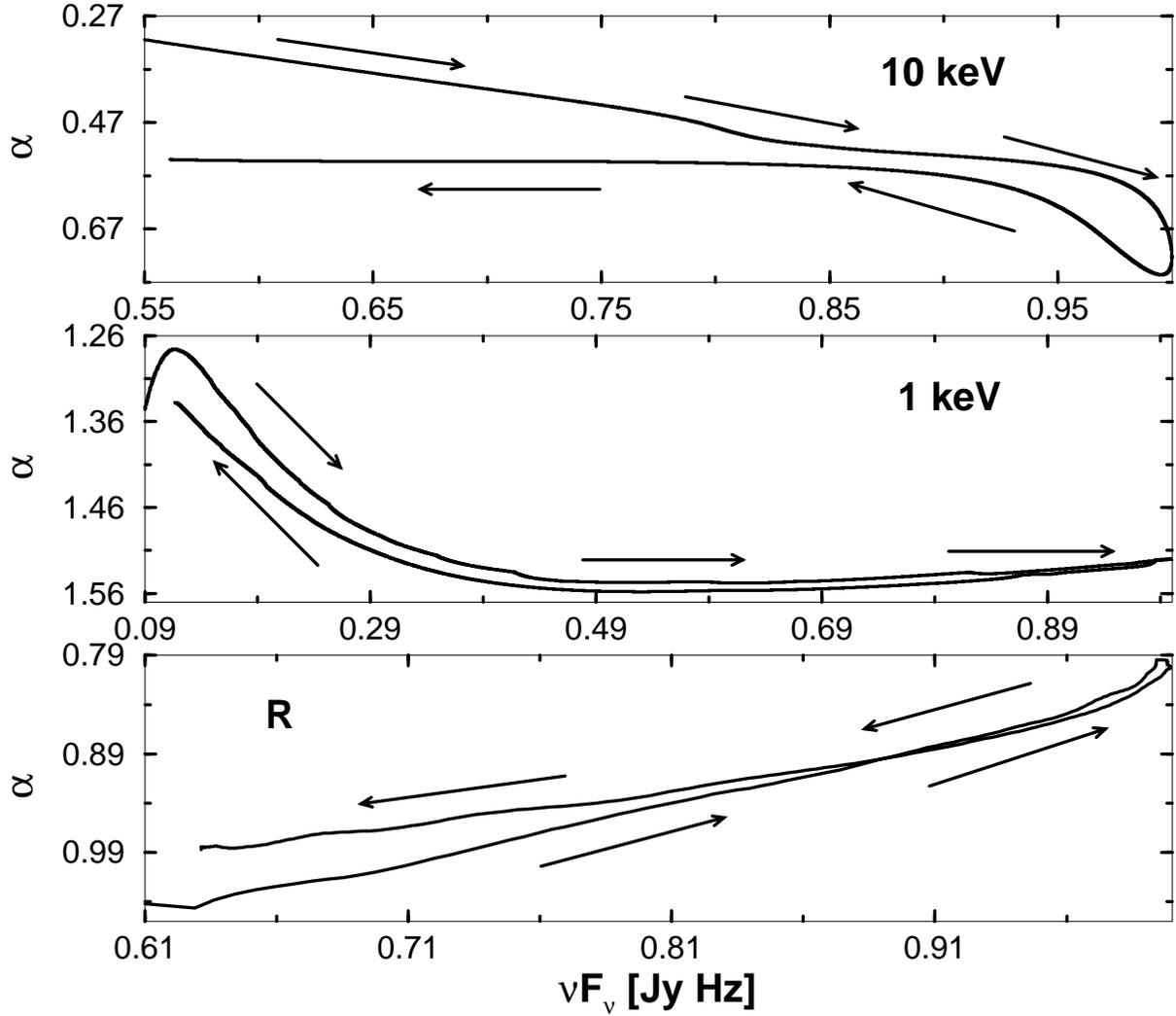}
\caption{Simulated spectral hysteresis pattern in the R-band, 1 keV and 10 keV
 energy regimes, shown in the three panels respectively. As can be seen, the 
hysteresis pattern starts to show up in the 10 keV energy regime.}
\label{hyst_combined1}
\end{figure}

\begin{figure}
\plotone{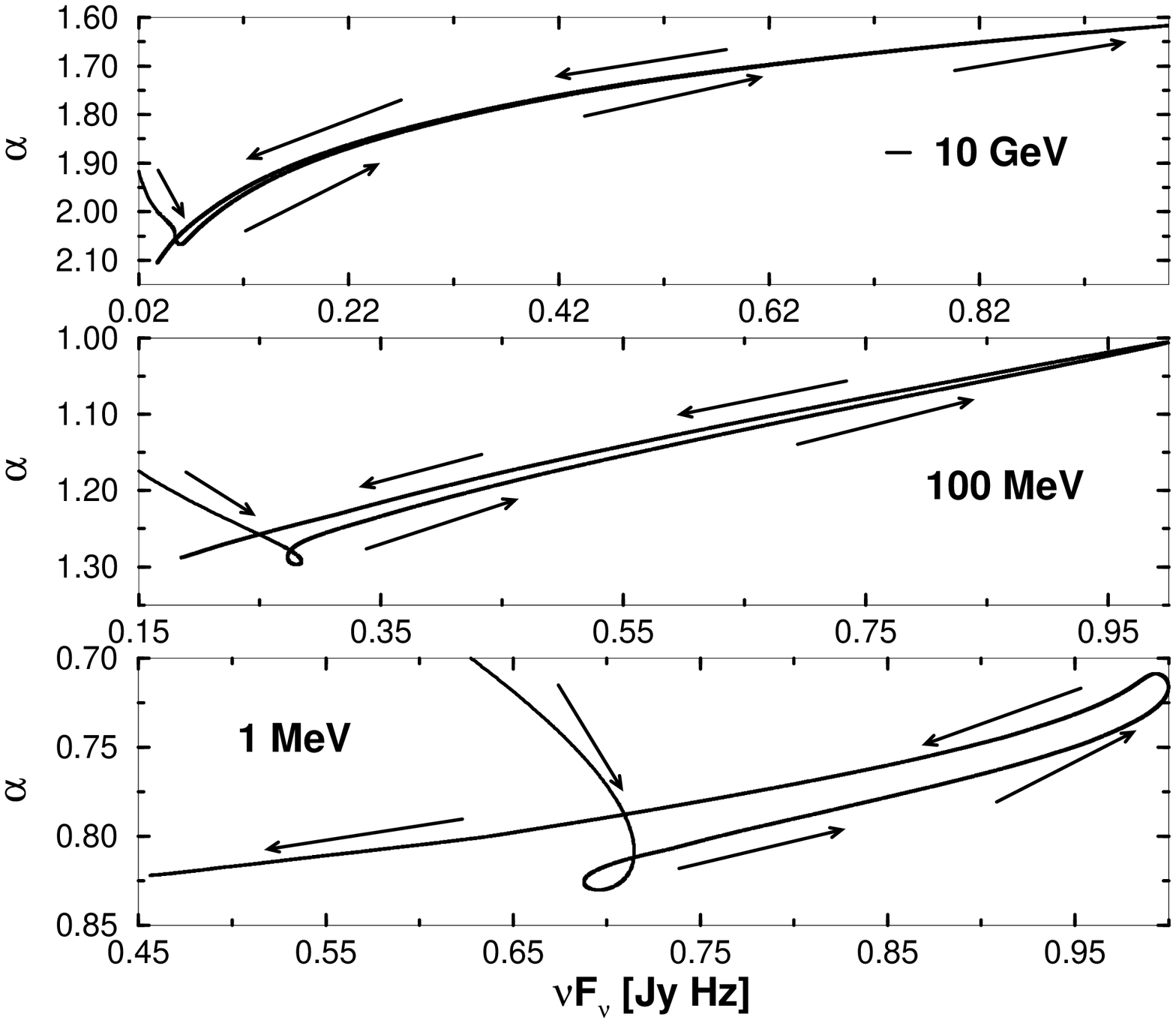}
\caption{Simulated hysteresis pattern for 1 MeV, 100 MeV and 10 GeV energy 
regimes, shown in the three panels repectively. The hysteresis pattern is
prominent for the 1 MeV energy regime but starts to become absent at higher
energies.} 
\label{hyst_combined2}
\end{figure}

As can be seen in Figure \ref{hyst_combined1}, spectral hysteresis patterns 
are not predicted for optical as well as soft X-ray photons. This is expected 
because the cooling timescale of their parent electron population is so short 
that what is observed is the average effect of this cooling over the dynamical 
timescale and hence any hysteresis pattern gets smeared out. On the other hand,
 one expects to see these patterns at higher energies because as explained 
earlier, this photon population comes from Compton upscattering off low-energy
 electrons, which have a longer cooling timescale and as a result the photon 
population gradually builds up over time and then dies away giving rise to a 
hysteresis pattern (see Figure \ref{hyst_combined2}). The slight spectral 
softening at 10 keV seen in its hysteresis pattern (see Figure 
\ref{hyst_combined1}) for higher values of $\nu F_{\nu}$ indicates a small 
synchrotron contribution near the peak of the flare.

The simulated instantaneous SED, for a pure SSC model, shows a definite 
presence of $\gamma$-ray emission in 3C66A, in the quiescent as well as the 
flaring state (see Figure \ref{sed_plot190} and \ref{sed_plot344}). The 
intrinsic cutoff of VHE emission in the flaring state, according to the 
simulations, for the time averaged spectrum is $\sim 1.0 \times 10^{24}$~Hz or
 4 GeV. In our simulations, the emission of VHE $\gamma$-ray photons is 
produced by the SSC mechanism in the quiescent as well as the flaring state. 
Figure \ref{sed_344_lightcurve} shows the simulated lightcurves for VHE 
photons and as can be seen the $\nu F_{\nu}$ value changes by 
$\sim 4.17 \times 10^{12}$~Jy Hz at 100 MeV. The variability in VHE photons is
 expected as they are the result of Compton upscattering off the higher energy 
electrons and due to this the hysteresis pattern is not seen at such high 
energies as the cooling timescale of such high energy electrons is very short 
(see Figure \ref{hyst_combined2}).

\begin{figure}
\plotone{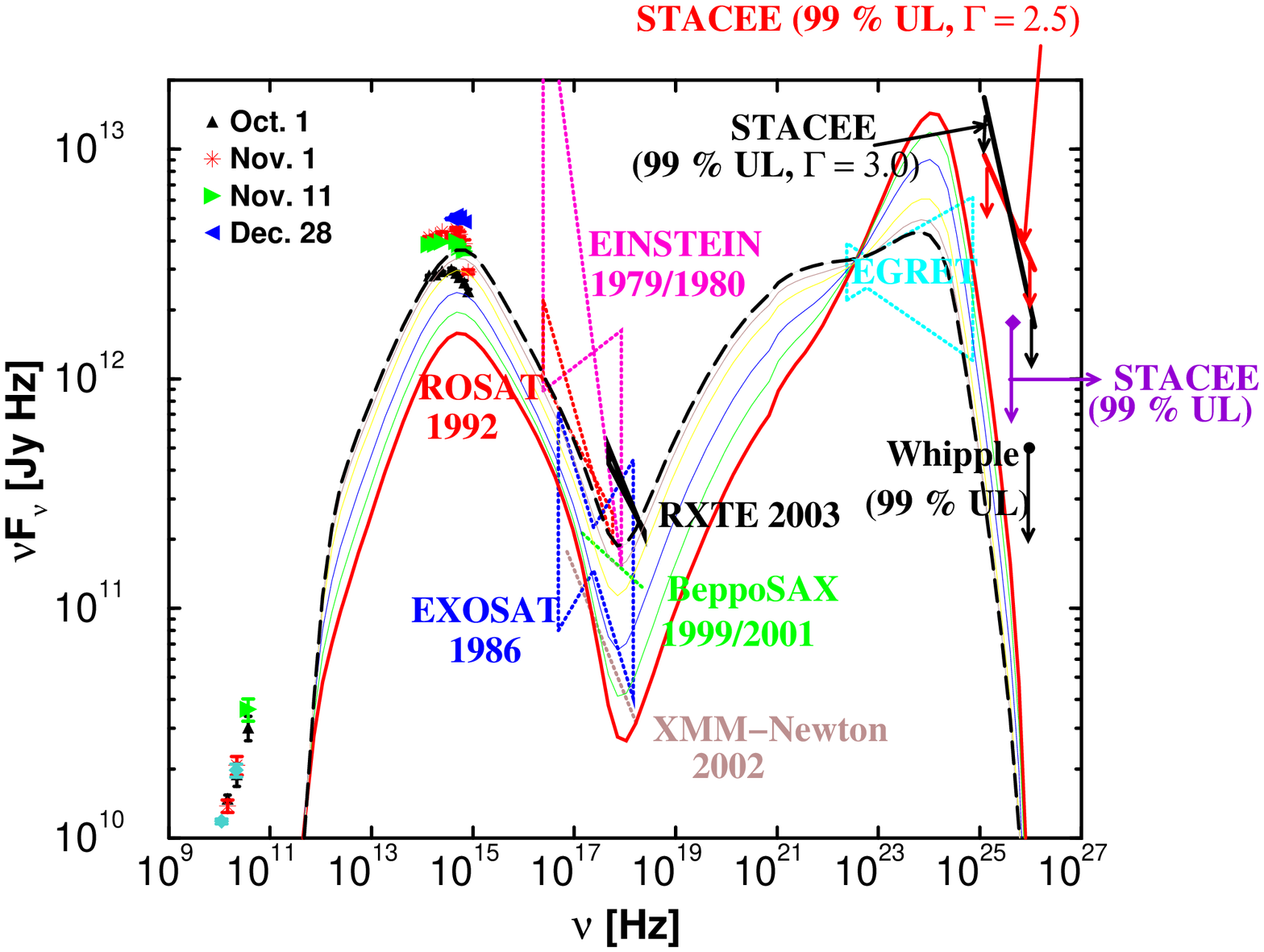}
\caption{Simulation of the effect of the BLR on the instantaneous spectral 
energy distribution of 3C~66A for the first 3 days of a simulation similar to 
Figure \ref{sed_plot344}. The curves in the figure denote the instantaneous 
spectra obtained from the simulation. The gray (red in the online version) 
solid line denotes one of the initial instantaneous spectrum at the 
beginning of the simulation whereas the black long-dashed line indicates the 
last spectrum obtained from the simulation.}
\label{sed_plot348}
\end{figure}

From Figure \ref{sed_plot348}, it can be seen that the high-energy component of
 3C~66A, in the flaring state, could start out with a dominant contribution of
 the EC emission, shown by the red solid line. But as the blob travels further
 away and passes the outer edge of the broad line region, the EC contribution 
becomes less significant and the SSC emission takes over. This is indicated by
 the black long-dashed line in the figure. We might actually find that this 
maximum contribution would be just enough to explain the historical EGRET flux
 and that there could be GeV flaring due to early external Comptonization.  

The effect of an optical depth due to the IIRB on the spectra of 3C~66A was 
also evaluated and was found to be insignificant in the energy range we are 
interested in as shown in Figure \ref{sed_timeav_344}. The optical depth due 
to the IIRB was determined using the analytic expression given in 
\cite{sm2006}. The $\gamma - \gamma$ absorption till $\sim 100$~GeV is 
negligible and becomes slightly observable at $\sim$ 200 GeV as the optical 
depth takes a value of, $\tau_{\gamma \gamma} \approx$ 2.9. Hence, the SSC 
emission cutoff value at $\sim$ 4 GeV is intrinsic.

\section{\label{summary}Summary}

An extensive analysis of the data of 3C~66A, obtained from the multiwavelength
 monitoring campaign on 3C~66A from July 2003 to April 2004, was carried out 
using a time-dependent leptonic jet model. The analysis was targeted towards 
understanding the dominant radiation mechanism in the production of the 
high-energy component of the SED of 3C~66A in the quiescent as well as the 
flaring state. Our simulations yielded predictions regarding the observable 
variability patterns in the X-ray as well as the VHE energy regimes where such 
patterns could not be detected during the campaign. The object was well 
sampled in the optical, especially in the R-band, during the campaign. It had 
exhibited several major outbursts ($\sim$10 days) in this regime with a 
varibility of $\Delta$m $\sim$ 0.3-0.5. The X-ray data covered the 3-10 
keV range with the onset of the high-energy component expected at $\geq$~10 
keV photon energies. Only upper limits in the VHE regime had been obtained.

The simulations from our model could successfully reproduce the observed SED 
as well as the optical spectral variability patterns. The model suggests the 
dominance of the SSC mechanism in the production of hard X-ray as well as VHE 
photons. On the other hand, soft X-ray photons exhibit spectral softening 
during flaring indicating the onset of the synchrotron component in this energy
 range. According to the simulated time-averaged spectrum, the synchrotron 
component is expected to cut off near 7 keV whereas the SSC component cuts off
 at $\sim$ 4 GeV.

A flaring profile that was Gaussian in time could successfully reproduce the 
observed flaring profile for a timescale of $\sim$ 10 days. The simulated 
varibility in R ($\Delta$m $\sim$ 0.55) agreed well with the observed 
variability. According to the simulations, the object flares up in R and B 
simultaneoulsy with $\tau^{\rm obs}_{\rm cool, syn}$ (37 minutes) being less 
than or equal to the light crossing time (2 hours) during flaring. No 
significant variability is predicted in the hard X-ray regime. This is due to 
the production of such photons from Compton upscattering off low-energy
 electrons with cooling timescales much longer than the light crossing time, 
$3R_{b}/4c$. On the other hand, the simulated lightcurves of VHE 
$\gamma$-ray photons exhibit significant variability as such photons are
produced from the Compton upscattering off higher energy electrons with 
shorter cooling timescales than the light crossing time. 

The effect of the optical depth due to $\gamma - \gamma$ absorption by the 
IIBR on the SED of 3C~66A was also evaluated. The simulations do not predict a
 significant effect on the SED due to the optical depth. The SSC emission 
cutoff predicted to be at $\sim$ 4 GeV can be taken as the intrinsic SSC 
emission cutoff value for this object. We predict the object to be well 
within the observational range of MAGIC, VERITAS and GLAST. Finally, the EC 
emission for this object was also calculated and it appears that the EC 
emission could be dominant in the high-energy component initially, but as the 
emission region travels further away from the BLR, the EC contribution becomes
 less significant and the SSC emission takes over. It is highly probable that 
this maximum contribution of the EC component might explain the historical 
EGRET flux and that there could be GeV flaring due to early external 
Comptonization.

\acknowledgments
This work was partially supported through NRL BAA 76-03-01, contract no. 
N00173-05-P-2004.

\end{document}